\documentstyle[jkas]{article}

\runningauthor {S.-H. KOOK ET AL.} \year{2010} \volume{43}
\beginpage{141}\endpage{152}
\runningtitle{NGC 4609 \& HOGG 15}

\begin{document}

\title{$UBVI$ CCD PHOTOMETRY OF THE OPEN CLUSTERS \\
NGC 4609 AND HOGG 15}

\author{Seung-Hwa Kook$^1$, Hwankyung Sung$^1$, and M. S. Bessell$^2$}
\address{$^1$ Department of Astronomy and Space Science, Sejong University,
	98 Kunja-dong, Kwangjin-gu, Seoul 143-747, Korea \\
	{\it E-mail: sungh@sejong.ac.kr}}

\address{$^2$ Research School of Astronomy and Astrophysics, Australian
	National University, MSO, Cotter Road, Weston, ACT 2611, Australia\\
	{\it E-mail: bessell@mso.anu.edu.au}}

\address{\normalsize{\it (Received July 12, 2010; Revised August 18, 2010; Accepted August 25, 2010)}}
\offprints{H. Sung}

\abstract{$UBVI$ CCD photometry is obtained for the open clusters NGC
4609 and Hogg 15 in Crux. For NGC 4609, CCD data are presented for the first 
time. From new photometry we derive the reddening, distance modulus
and age of each cluster - NGC 4609 : $E(B-V) =$ 0.37 $\pm$
0.03, $V_0 - M_V =$ 10.60 $\pm$ 0.08, $\log \tau=$ 7.7 $\pm$ 0.1; Hogg 15 :
$E(B-V) =$ 1.13 $\pm$ 0.11, $V_0 - M_V =$ 12.50 $\pm$ 0.15, $\log \tau 
\lesssim$ 6.6. The young age of Hogg 15 strongly implies that WR 47 is a member
of the cluster. We also determine the mass function of these clusters and
obtain a slope ($\Gamma = -1.2 \pm 0.3$) for NGC 4609 which is normal and 
a somewhat shallow slope ($\Gamma = -0.95 \pm 0.5$) for Hogg 15.
}


\maketitle

\section{INTRODUCTION}

Although massive stars play an important role in the chemical and dynamical
evolution of a galaxy, knowledge relating to their formation processes
and environments is still very scanty due to the rareness of
massive O, Wolf-Rayet (WR), and luminous blue variable (LBV) stars.
From the correlation between WR stars and the turn-off mass of young open
clusters in the Galaxy, Massey et al. (2001) found that early-type WN (WNE) 
stars appear to have evolved from a large range of high mass stars,
whereas WN7 or LBVs are only found in clusters with the highest
turn-off masses. 

WR 47 (= HDE 311884 - WN6 + O5:V) is located at the periphery of the compact open cluster 
Hogg 15. The relationship between WR 47 and Hogg 15 is a key to studying 
the evolutionary status and progenitor mass of WR 47. The double-line binary
system WR 47 has attracted many observations because the system consists of
very massive stars with a relatively high mass ratio ($q_{WR} \equiv {m_{\rm WR} \over {m_{\rm O}
+ m_{\rm WR}}}$ = 0.85) (van der Hucht 2001). Niemela et al. (1980) found
that the WN6 star in WR 47 is more massive than 40 M$_\odot$ from the orbit
of the double-line binary system. Lamontagne \& Moffat (1987) detected about
0.1 mag variability for WR 47, and showed that the variability was not caused
by the pulsation-like intrinsic variability of WR stars. Later, Moffat et al.
(1990) determined the mass of each component and the mass loss rate from
the revised ephemeris based on photometry and polarimetry. The masses they
determined were 48 $\pm$ 5 M$_\odot$ for WN6 and 57 $\pm$ 6 M$_\odot$ for O5:V.
Lamontagne et al. (1996) revised the mass of each component (51 M$_\odot$ for 
WN6 and 60M$_\odot$ for O5:V). The mass of the companion is somewhat larger than
its spectroscopic or evolutionary mass ($\sim$ 40M$_\odot$). The mass
of each component is strongly dependent on the geometry of the system,
in particular, the adopted inclination angle.

Feinstein \& Marraco (1971) determined the distance ($d = 1320$ pc) and age
($6 \times 10^7$ yrs) of NGC 4609, and found a somewhat large spread around
its main sequence. Muzzio et al. (1976) and Lod\'en (1979) studied this region
as part of a project studying the loose clustering of young stars
or the spiral arm structure of the Galaxy. 
Because NGC 4609 is an intermediate-age open cluster at a relatively large 
distance with few members, it did not attract modern CCD observation until now.
Moffat (1974) performed $UBV$ 
photoelectric photometry of 23 stars in Hogg 15, and determined the distance 
($d = 4.2 \pm 0.4$ kpc) and age ($\lesssim 8 \times 10^6$ yrs) of the cluster.
He concluded that the WN6 star HDE 311884 is likely to be a member of Hogg 15.

Orsatti et al. (1998) performed polarimetry on 23 stars in Hogg 15. 
They established the existence of intracluster dust associated with Hogg 15
with a slightly different grain size distribution. They also showed WR 47 to 
be highly polarized with a strong time variation.
Sagar et al. (2001) and Piatti \& Clari\'a (2001) performed CCD photometry for
the compact open cluster Hogg 15. Interestingly they obtained very discrepant
results. Sagar et al. (2001) derived the distance
and age of Hogg 15 as 3.0 $\pm$ 0.3 kpc and 6 $\pm$ 2 Myr, respectively.
While Piatti \& Clari\'a (2001) determined 
2.6 $\pm$ 0.08 kpc and $\sim$ 300 Myr, respectively. From their relatively
large age for Hogg 15,  Piatti \& Clari\'a (2001) ruled out WR 47 being a 
cluster member. One aim of our work is to disentangle the discrepancy 
between the results of Sagar et al. (2001) and Piatti \& Clari\'a (2001).

In Section 2 we present the $UBVI$ photometric data for 7208 stars. We compare
our data with those of previous studies. The reddening law of the observed
region is derived in Section 3. From the surface density variation of candidate
cluster stars we determine the radius of NGC 4609 and Hogg 15. The photometric 
members for both clusters are selected in the same section. The age and
mass function are derived in Section 4. The discussion and summary are presented
in Section 5 and Section 6 respectively.

\section{OBSERVATIONS AND DATA REDUCTION}

\begin{figure}[h]
\begin{center}
\epsfxsize=8.5cm \epsfbox{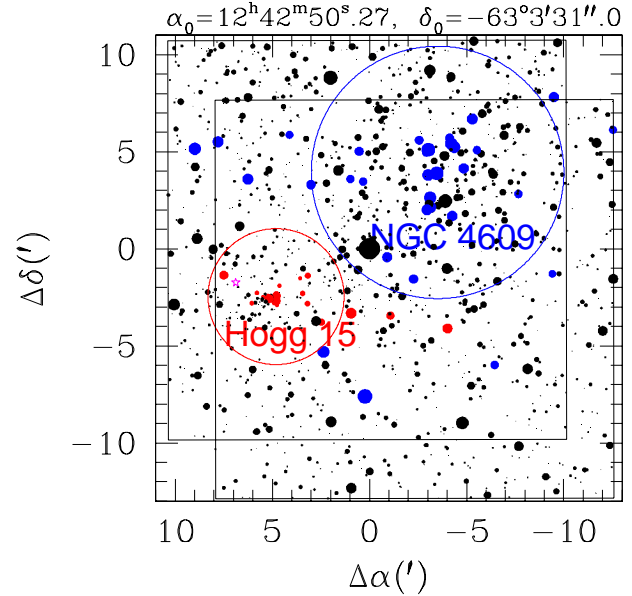}
\end{center}
\caption{Finding chart, centered at HD 110432 ($\alpha$ = 12$^h$ 42$^m$ 
20.$^s$3, $\delta$ = -63$^\circ$ 3$'$ 31.$''$0, J2000.0). The size of 
the symbol is proportional to the magnitude of the star. Blue and red-colored 
symbols represent less reddened nearby ($E(B-V) =$ 0.3 -- 0.45 \& $d \approx 
1.3$ kpc) and more highly reddened distant ($E(B-V) =$ 1.0 -- 1.5 mag \& $d 
\approx 3.2$ kpc) stars, respectively. The circles represent
the radius of NGC 4609 and Hogg 15, respectively.
Two large squares show the field of view (FOV) of the two observations. The square 
slightly shifted to the northeast is the FOV of the 1997 March 2 observation
($UBV$ only) and that slightly shifted to the southwest is the FOV of the 1997 
March 4 observation ($UBVI$).  \label{map}}
\end{figure}

{\scriptsize
\begin{deluxetable}{@{}rccccccc@{}c@{}c@{}c@{}cc@{}cl@{}}
\setlength{\tabcolsep}{-1mm}
\tablecaption{Photometric Data\tablenotemark{a}}
\tablehead{
\colhead{ID} & \colhead{$\alpha (J2000.0)$} & \colhead{$\delta (J2000.0)$} & 
\colhead{$V$} & \colhead{$V-I$} & \colhead{$B-V$} & \colhead{$U-B$} &
\colhead{$\epsilon_{V}$} & \colhead{$\epsilon_{V-I}$} & 
\colhead{$\epsilon_{B-V}$} & \colhead{$\epsilon_{U-B}$} &
\colhead{n$_{obs}$} & \colhead{2MASS} & \colhead{Sp type} & \colhead{Remark}
}
\startdata
1\tablenotemark{b}&12:42:50.27&-63:03:31.0& 5.316&\nodata& 0.248&-0.813&0.052&\nodata&0.018&0.027&7 0 7 6&12425028-6303310 &B2pe &HD 110432 \\
 3902&12:42:52.28&-63:11:07.3& 9.024& 0.410& 0.239&-0.348&0.002&0.005&0.004&0.004&1 1 1 1&12425228-6311072 &B2.5IV-V&HD 110433 \\         
 4573&12:43:07.85&-62:54:42.1& 9.347&  --  & 1.455& 1.476&0.002&  -- &0.006&0.006&1 0 1 2&12430790-6254425 &K0.5III&HD 110478 \\
 2331&12:42:15.81&-63:01:03.1& 9.555& 0.768& 0.709& 0.362&0.002&0.006&0.004&0.004&2 1 1 1&12421586-6301035&        &          \\
  803&11:02:48.36&-61:13:51.6& 9.812& 0.128& 0.098&-0.507&0.014&0.027&0.016&0.014&1 1 1 1&11024831-6113512 &       &          \\
 2668&12:42:23.60&-62:58:24.7& 9.577& 0.280& 0.149&-0.438&0.004&0.005&0.005&0.003&2 2 1 1&12422361-6258249&B8V     &HD 110373 \\
 2506&12:42:19.74&-62:59:37.5& 9.605& 0.308& 0.188&-0.359&0.003&0.003&0.004&0.004&1 1 1 1&12421975-6259376 &       &          \\
 6544&12:43:51.03&-63:05:14.9&10.757& 1.437& 0.837&-0.190&0.004&0.005&0.004&0.015&2 2 2 2&12435102-6305148&WN6+O5V & HDE311884   \\
 5599&12:43:29.91&-62:57:36.7&14.224& 5.803& 2.177& 0.568&0.004&0.008&0.011&0.013&3 1 3 2&12432995-6257369& Mira   &              \\
\enddata
\tablenotetext{a}{This is a sample of the full table, which is available from HS.}
\tablenotetext{b}{Photometric data from the SIMBAD database (http://simbad.u-strasbg.fr/simbad/)}
\label{tab1}
\end{deluxetable}
}

$UBVI$ CCD photometry for NGC 4609 and the Hogg 15 region was performed during
two nights in 1997 March using the 1-m telescope at Siding Spring Observatory. 
On March 2, due to limited observing time and the night not being 
photometric, we could observe the region only in $UBV$. We therefore observed
the region again on March 4 in $UBVI$.
We obtained images for two sets of exposure times: long, $2 \times 30s$ in $I$,
$2 \times 60s$ in $V$, $2 \times 100s$ in $B$, and $2 \times 300s$ in $U$;
and short, $5s$ in $V$ and $I$, $10s$ in $B$, and $30s$ in $U$.
We also observed two Landolt standard 
regions (SA 98 \& PG 1312-086) on March 4. The standard transformation 
relations in Sung, Bessell \& Lee (1998) are used for $UBV$, but for the $I$ 
transformation we used the relation in Sung et al. (2008).
The atmospheric extinction coefficients on March 4 were 0.094 ($\pm$ 0.011),
0.190 ($\pm$ 0.013), 0.355 ($\pm$ 0.008), and 0.580 ($\pm$ 0.019) in $I$,
$V$, $B$, and $U$, respectively. The secondary extinction coefficients
in $B$ and $U$ from Sung et al. (1998) have been adopted and used.

\begin{figure}[!h]
\begin{center}
\epsfxsize=7.0cm \epsfbox{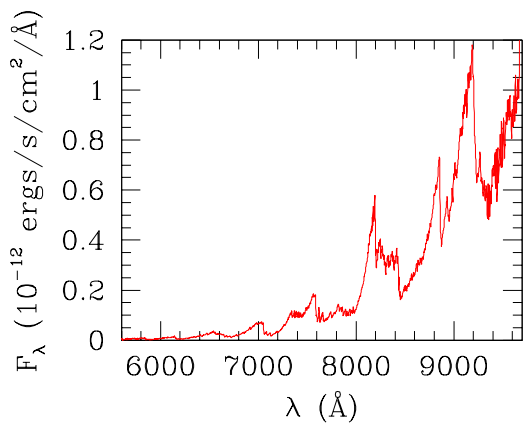}
\end{center}
\caption{Spectrum of ID 5599 obtained with WiFeS on the 2.3m telescope at
Siding Spring Observatory.
\label{id5599}}
\end{figure}

Instrumental magnitudes were obtained using IRAF/ DAOPHOT using point
spread function fitting. All the instrumental magnitudes were transformed
to the standard $UBVI$ system using Landolt (1992) standard stars observed on
1997 March 4. The photometric data for 7208 stars were obtained and a sample
of the data is presented in Table \ref{tab1}. Fig. \ref{map} is the finding 
chart for stars brighter than $V = 17$ mag. The only saturated star 
was the peculiar Be star HD 110432, which shows an enormously high X-ray 
temperature (see Smith \& Balona 2006). The data for HD 110432 were obtained 
from the SIMBAD database. We have also identified 2MASS counterparts of 
the optical sources. The star ID 5599 is very red in the optical as well as 
in the near-IR 2MASS $JHK_s$. A spectrum of the star obtained with WiFeS 
on the 2.3m telescope at SSO (see Fig. \ref{id5599})
on 2010 July 19 shows strong TiO and VO bands. The star is clearly a Mira.
 
\begin{figure}[!h]
\begin{center}
\epsfxsize=7.0cm \epsfbox{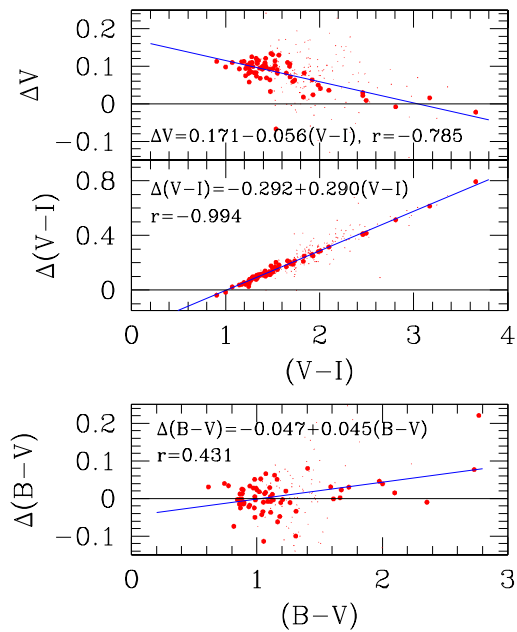}
\end{center}
\caption{Comparison of our photometry with that of Piatti \& Clari\'a (2001).
Large and small dots represent bright ($V \leq 17$ mag) and faint stars
($V > 17$ mag), respectively. The regressions were performed only for bright 
stars. There are large systematic differences in $V$ and $V-I$ between our 
data and those of Piatti \& Clari\'a (2001), but the difference in $B-V$
is not as evident.  \label{piatti}}
\end{figure}

The $UBVI$ magnitudes and colors from our study are compared with the previous
photoelectric or CCD photometry of Feinstein \& Marraco (1971), Moffat (1974),
Muzzio et al. (1976), Lod\'en (1979), Piatti \& Clari\'a (1981), and Sagar
et al. (2001) in Table \ref{tab2}. The meaning of $\Delta$ is in the sense of
our data minus that of the others. Piatti \& Clari\'a (2001) observed
many standard regions from Landolt (1992), but the others, except Moffat 
(1974)\footnote{No information on the standard stars they observed 
is available.},
used SAAO E-regions as standards. Although our system is tightly 
matched to the Landolt system, the difference between our data and others 
is very small in $B-V$. The difference in $U-B$ should be more negative 
as the $U-B$ scale of the Landolt system is bluer for blue stars (Menzies et 
al. 1991; Lim et al. 2009). The data by Sagar et al. (2001) do not show 
such a trend although they observed SAAO E-regions as standard stars.
The difference in $V$ or $V-I$ of Piatti \& Clari\'a (2001) is very large
and systematic (see Fig. \ref{piatti}). 
The source of such a systematic difference cannot be traced 
back because they did not provide the atmospheric extinction coefficients and
transformation coefficients. The difference in $B-V$ is, on the other hand,
not as systematic (see
Table \ref{tab2}). The $V$ and $V-I$ scales of Sagar et al. (2001) do not show 
any systematic differences, but do show somewhat large shifts in zero points.

{\scriptsize
\begin{deluxetable}{lcccccccc}
\tablecolumns{8}
\tablecaption{Comparison with Other Photometry \label{tab2} }
\tablewidth{0pt}
\tablehead{
\colhead{Author} & \colhead{$\Delta V$} &
\colhead{n (n$_{\rm ex}$)\tablenotemark{a}} & \colhead{$\Delta (B-V)$} &
\colhead{n (n$_{\rm ex}$)\tablenotemark{a}} & \colhead{$\Delta (U-B)$} &
\colhead{n (n$_{\rm ex}$)\tablenotemark{a}} & \colhead{$\Delta (V-I)$} &
\colhead{n (n$_{\rm ex}$)\tablenotemark{a}} }

\startdata
Feinstein \& Marraco (1971) & +0.032 $\pm$ 0.047 &58 (8 ) &
-0.001 $\pm$ 0.027 & 60 (6) & -0.046 $\pm$ 0.058 & 57 (8) &
\nodata            &        \\
Moffat (1974)      & +0.123 $\pm$ 0.225 & 22 (1) &
-0.007 $\pm$ 0.063 & 21 (2) & -0.005 $\pm$ 0.126 & 20 (3) &
\nodata            &        \\
Muzzio et al.(1976)& +0.026 $\pm$ 0.122 & 6 (1) &
+0.009 $\pm$ 0.017 & 6 (1) & -0.081 $\pm$ 0.041 & 6 (1) &
\nodata            &        \\
Lod\'en (1979)     & -0.008 $\pm$ 0.019 & 17 (0) &
+0.003 $\pm$ 0.014 & 17 (0) & -0.027 $\pm$ 0.028 & 16 (1) &
\nodata            &        \\
Piatti \& Clari\'a (2001)\tablenotemark{b} & see Fig. \ref{piatti} & 67 (1) & +0.006 $\pm$ 0.046 & 
67 (1) & \nodata &  & see Fig. \ref{piatti} & 67 (1) \\
Sagar et al.(2001)\tablenotemark{b} & +0.105 $\pm$ 0.022 & 46 (8) & 
-0.034 $\pm$ 0.023 & 47 (4) & -0.006 $\pm$ 0.063 & 37 (2) &
+0.066 $\pm$ 0.027 & 51 (3) \\
\enddata
\tablenotetext{a}{The number of stars excluded in the comparison.}
\tablenotetext{b}{Comparisons were made only for $V \leq 17$ mag.}
\end{deluxetable}
}

\section{PHOTOMETRIC DIAGRAMS}

\subsection{Reddening Law}

\begin{figure}[!b]
\begin{center}
\epsfxsize=7.0cm \epsfbox{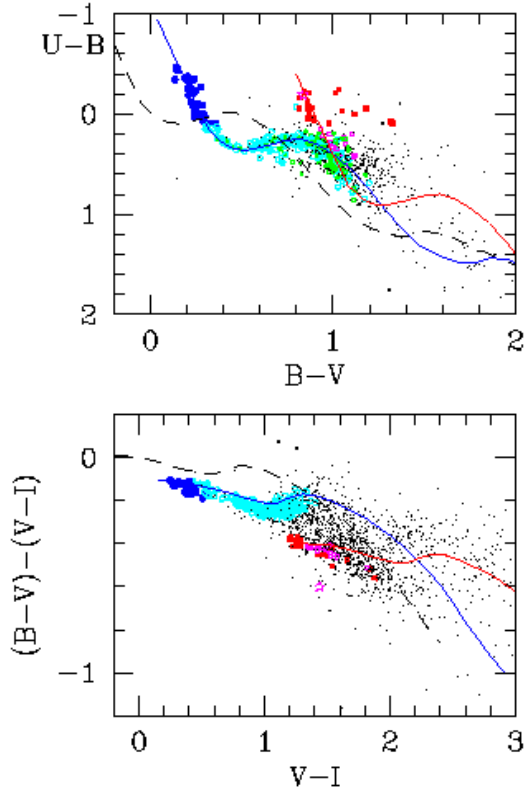}
\end{center}
\caption{The ($U-B$, $B-V$) and ($B-V$, $V-I$) color-color diagrams. 
The thin dashed line represents the ZAMS 
relation, while the two solid lines represent the ZAMS relations 
reddened by $E(B-V)$ = 0.37 (blue) and 1.13 (red), respectively. (Blue) Filled 
circles and (red) squares are the stars belonging to these two groups, while
open symbols are the photometric members of each group (see Section 3.3 for the
selection criteria of photometric members). The smallest (green) circle 
represents the photometric members selected only from the ($V$, $B-V$)
diagram. The star symbol (magenta)
denotes WR 47. Small dots and crosses represent, respectively, 
good data ($\epsilon < 0.1$) and bad data ($\epsilon \geq 0.1$). \label{ccd} }
\end{figure}

Correction for interstellar reddening is the first step in determining
the distance and age of open clusters. Fig. \ref{ccd} shows the color-color
diagrams of the observed region. Evidently there are at least two groups of
reddened blue stars. The less reddened blue stars are the B type stars belonging
to or around the intermediate-age open cluster NGC 4609, while the more highly reddened 
group of stars are stars in and around the young open cluster Hogg 15.
Most blue stars in NGC 4609 show little scatter in reddening, but those
belonging to Hogg 15 show a large scatter. Unlike WR 37 in the Ruprecht
93 region (Cheon et al. 2010), the $UBV$ colors of WR 47 is very similar
to those of normal early type stars in Hogg 15. The mean $E(B-V)$ of 
NGC 4609 and Hogg 15 is 0.37 ($\pm$ 0.03) mag and 1.17 ($\pm$ 0.11) mag, 
respectively. The median $E(B-V)$ of Hogg 15 is 1.13 mag. The spatial 
distribution of the highly reddened group in Fig. \ref{map} is not circular 
but elliptical in shape.

There are three reddened blue stars between the two reddened zero-age main
sequence (ZAMS) relations.
The one on the upper left part of the ($U-B$, $B-V$) diagram is HD 110432,
the brightest star in the observed region. Although the star is in front
of NGC 4609 ($\pi_{\rm HD ~110432} = 3.32 \pm 0.56$ mas), the reddening 
$E(B-V)$ is 0.58 mag. This means that more than 0.28 mag in $E(B-V)$ (i.e.
$E(B-V)_{\rm HD~110432}$ - $E(B-V)_{\rm NGC~4609}^{min}$ = 0.58 - 0.30 = 
0.28) is caused by the circumstellar material around HD 110432.
The other two stars between the two reddened ZAMS lines are ID 3521 and ID 3585.
The distance modulus from ZAMS fitting is between 11.0 mag and 11.7 mag which 
is very similar to the distance modulus of the Cen OB1 association (Humphreys
1978). A few stars in the observed region may therefore be young stars
belonging to the Cen OB 1 association.

\begin{figure}[!h]
\begin{center}
\epsfxsize=8.5cm \epsfbox{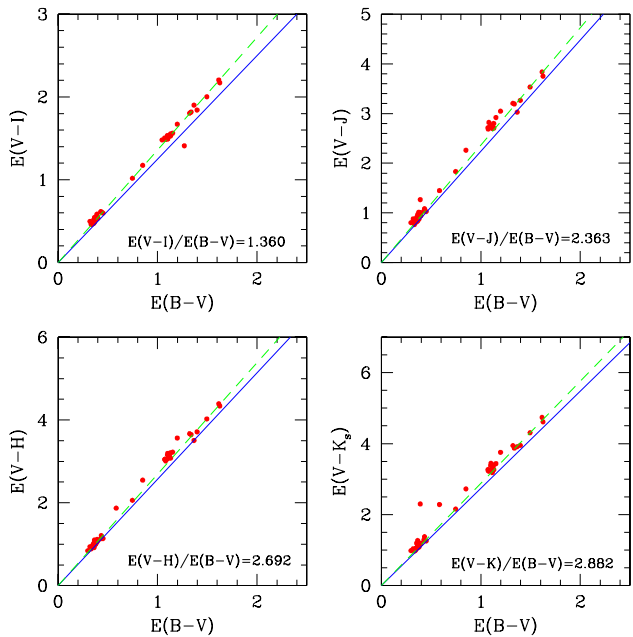}
\end{center}
\caption{Color excess ratios. The color excess of each color is calculated
using the relation between intrinsic colors. The thick solid line represents
the color excess relation for the normal reddening law, i.e. $R_V$ = 3.1,
while the dashed line denotes the color excess ratio for the observed region
($R_V$ = 3.2).  \label{red_law} }
\end{figure}

Knowledge about the reddening law, especially the total-to-selective 
extinction ratio, $R_V \equiv A_V / E(B-V)$, is very important in estimating
the distance to an object. We have employed the same method as used in the Ruprecht
93 region (Cheon et al. 2010). The reddening $E(B-V)$ of individual 
early type stars is calculated from the color-color diagram in Fig. 
\ref{ccd}. The color excess in each color is calculated using the relation
between intrinsic colors.

Fig. \ref{red_law} shows the relation between $E(B-V)$ and color excesses in
other colors. The thick solid line represents the color excess relation for the
normal case, i.e. $R_V = 3.1$. Evidently early type stars in the observed region
are slightly elevated above the normal values. The median value of $E(V-I) / E(B-V)$, 
$E(V-J) / E(B-V)$, $E(V-H) / E(B-V)$, and $E(V-K_s) / E(B-V)$ for $E(B-V)
\geq$ 1.3 mag is 1.360 $\pm$ 0.024, 2.363 $\pm$ 0.067, 2.692 $\pm$ 0.066, 
and 2.882 $\pm$ 0.054, respectively. If we adopt the relation between 
the color excess ratio and $R_V$ of Guetter \& Vrba (1989), the color excess 
ratios give $R_{V} = 3.18 \pm 0.06$. We adopt $R_V = 3.2$ for the observed
region. Several blue stars in Hogg 15 ($E(B-V) \approx$ 1.1 mag) show slightly 
higher values in $E(V-J)$, $E(V-H)$, and $E(V-K_s)$, but normal values in
$E(V-I)$. This may be caused by scattered light or the lower angular resolution 
of the telescope used in 2MASS. But we cannot rule out the possibility
that the $R_V$ at the center of Hogg 15 is slightly larger than normal 
(see discussion in Section 3.3.3 or Orsati et al. 1998).

\subsection{\rm Surface Density Variation and Radius}

\begin{figure}[!h]
\begin{center}
\epsfxsize=8.5cm \epsfbox{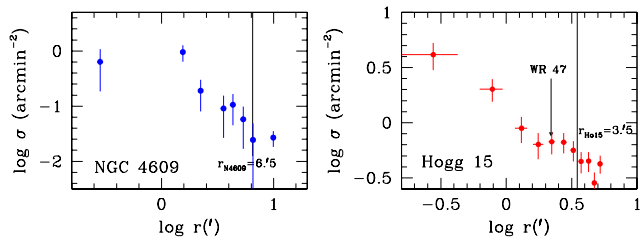}
\end{center}
\caption{The surface density variation of blue stars in NGC 4609 (left)
and of stars near the compact young cluster Hogg 15 (right).  \label{rho} }
\end{figure}

As there is a spiral arm at about 1 kpc in the direction of NGC 4609 (see
Fig. 5 of Hou et al. 2009) and the distance to NGC 4609 is about 1.3 kpc,
it is reasonable to expect contamination from spiral arm field stars 
in the color-magnitude diagram (CMD) of NGC 4609. Since there is no
proper motion or other membership study for NGC 4609 and Hogg 15, 
surface density variation is the only way to determine 
the radius of each cluster.\footnote{We have checked proper motion data in the 
Tycho catalogue, but could not find any concentration of candidate cluster
stars in the ($\mu_\alpha$, $\mu_\delta$) plane probably because the proper
motion of cluster stars is similar to that of field stars in the Sgr-Car arm.}

For NGC 4609 we have calculated the surface density of blue stars in annuli
from the apparent center of NGC 4609 ($\alpha$ = 12$^h$ 42$^m$ 19.$^s4$,
$\delta$ = -62$^\circ$ 59$'$ 35$''$). The left panel of Fig. 
\ref{rho} shows the radial variation of the surface density of blue stars.
The surface density of blue stars decreases rapidly and reaches
the surface density of the field at $r \approx 6.'5$. 
We tentatively adopt the radius of NGC 4609 as 6.$'$5.

Hogg 15 is a small compact open cluster and the number of early type members 
is therefore very small. To determine the radius of Hogg 15, we have
calculated the surface density of stars with $V$ = 10 -- 17 mag and $B-V$ =
0.7 -- 1.4 mag from the centroid of early type member stars (red dots in Fig.
\ref{map}). We have adopted the center of Hogg 15 to be $\alpha$ = 12$^h$
43$^m$ 32.$^s$8, $\delta$ = -63$^\circ$ 5$'$ 59$''$.
The variation of surface density is shown in the right panel of Fig.
\ref{rho}. The surface density decreases rapidly up to $r \approx 1.'7$,
shows a plateau between $r$ = 1.$'$7 -- 3.$'$0, and decreases to the
background level at $r$ = 3.$'$5. WR 47 is about 2.$'$2 from the adopted
center. From surface density variations, Hogg 15 is a cluster with a core
and a halo surrounding the core. WR 47 is in the halo of Hogg 15.

\subsection{Photometric Members and Distance}

\begin{figure}[!t]
\begin{center}
\epsfxsize=8.50cm \epsfbox{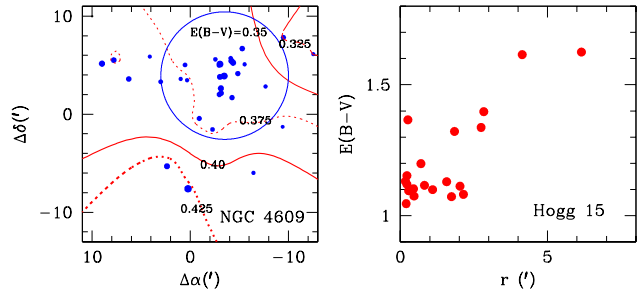}
\end{center}
\caption{Reddening map of NGC 4609 and the variation of $E(B-V)$ within Hogg 15.
\label{redmap} }
\end{figure}

\subsubsection{Membership Selection Criteria}

Sung \& Bessell (1999) have devised photometric membership selection criteria.
The method makes use of a merit of multicolor photometry - the different
ZAMS relations for different colors and the different effects of reddening on 
different colors.  The method involves first calculating the distance moduli
both in the [$V_0$, $(B-V)_0$] and [$V_0$, $(V-I)_0$] diagrams
after correcting for the reddening of the star. Then, one calculates
the average and differences in the distance moduli for a given star.
Ideally, members of a cluster should have the same distance moduli. In actual
fact, due to photometric errors the distance moduli differ slightly.
In addition, metallicity (in the case of M35 - Sung \& Bessell 1999) or 
chromospheric activity (in the case of NGC 2516 - Sung et al. 2002) also
affects the differences in distance moduli because they have different
effects on different colors. We select the members of a cluster if (1) 
the average value of the distance modulus is between $(V_0 - M_V)_{cl} - 0.75 
- 2 \sigma_{V_0 - M_V}$ and $(V_0 - M_V)_{cl} + 2 \sigma_{V_0 - M_V}$ and 
(2) the difference in the distance moduli is less than $\pm 2.5 \sigma_{V_0 
- M_V}$. The factor -0.75 mag is added to take into account the effect of 
equal mass binaries. We adopt the distance moduli of NGC 4609 and
Hogg 15 as 10.60 ($\pm$ 0.08) mag and 12.5 ($\pm$ 0.15) mag, respectively.

This method only works well when photometry is very accurate on an absolute 
scale. If the cluster is distant or if there are many field stars with similar 
reddenings and distances, the usefulness of this method is very limited. In 
such cases we should first correct for the contribution of field interlopers. 
To estimate the distance modulus of a star we need to know the reddening. 
We derive the reddening map of NGC 4609 using the less reddened blue stars 
and it is shown in Fig. \ref{redmap}. The reddening map of NGC 4609 is very 
smooth. We estimate the reddening of faint stars by interpolating the reddening 
map. We select photometric members of NGC 4609 and Hogg 15 down to $V$ = 17 
mag. The stars in the region observed only on 1997 March 2 (northern edge and 
eastern edge of Fig. \ref{map}) have $UBV$ data only, and therefore the same
photometric membership selection criteria cannot be applied. The photometric
members of these regions have been selected only from the [$V_0$, $(B-V)_0$]. 
These stars
are marked as the smallest circles in Fig. \ref{ccd} or \ref{cmd}.

\subsubsection{NGC 4609}

\begin{figure}[!h]
\begin{center}
\epsfxsize=7.0cm \epsfbox{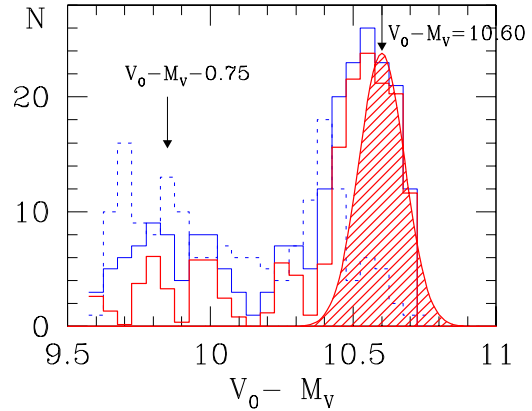}
\end{center}
\caption{The distribution of distance moduli of stars in and around NGC 4609.
The thin solid and dotted histograms represent the distribution of distance 
moduli of photometric members within and outside NGC 4609, respectively.
The thick solid histogram shows the distribution of distance moduli
subtracted for the contribution of field stars. The hatched area shows 
a gaussian function fit to the distribution of the right hand portion 
which approximates the distance modulus distribution of single stars.
\label{dist} }
\end{figure}

Fig. \ref{dist} shows the distribution of distance moduli of photometric
members in NGC 4609. The thin solid histogram represents the distribution of
distance moduli of photometric members within the cluster radius, while
the dotted histogram shows the distribution of distance moduli of stars 
outside the cluster radius which meet the same photometric membership 
selection criteria. The thick solid
histogram represents the distribution of distance moduli subtracted for
the contribution of field stars. The dotted histogram in Fig. \ref{dist}
shows a slight shift to the smaller distance modulus relative to the thin
solid histogram. This fact implies that the field stars having similar
photometric characteristics to NGC 4609 are slightly brighter than
the stars in NGC 4609, which relates to the finding of Feinstein \& Marraco
(1971) that there is a large spread in the distance moduli of stars in NGC 4609.

\begin{figure}[!h]
\begin{center}
\epsfxsize=8.50cm \epsfbox{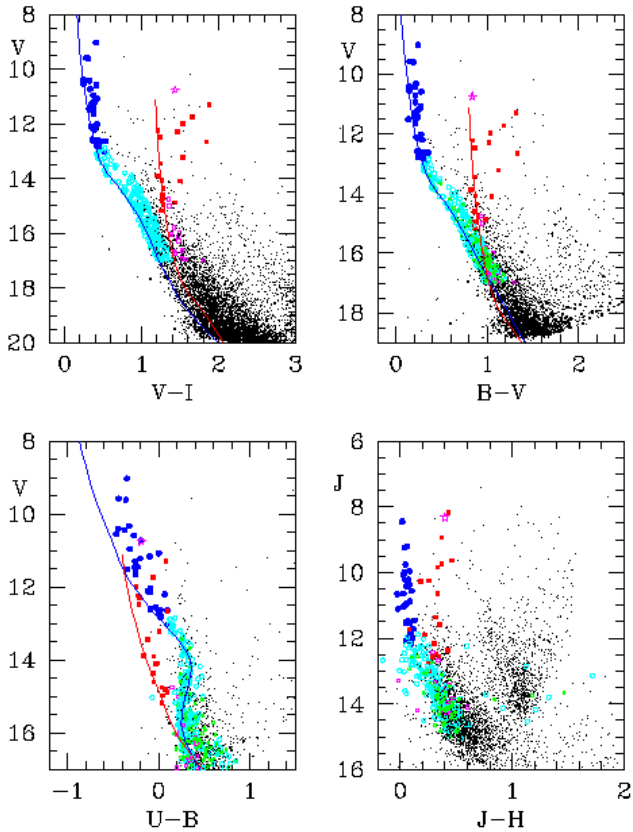}
\end{center}
\caption{Color-magnitude diagrams. Two solid lines represent the
ZAMS relations reddened by $E(B-V)$ = 0.37 and $V_0 - M_V = 10.6$, and
$E(B-V)$ = 1.13 and $V_0 - M_V = 12.5$, respectively. All the other symbols 
are the same as in Fig. \ref{ccd}. \label{cmd} }
\end{figure}

The shaded area in Fig. \ref{dist} represents a gaussian function fit
to the distance moduli distribution to estimate the distance modulus and
the binary fraction of NGC 4609. The distance modulus of NGC 4609 from 
the gaussian function fit is 10.60 $\pm$ 0.08 mag ($d$ = 1.32 $\pm$ 0.05 kpc). 
This value is the same as the distance determined by Feinstein \& Marraco
(1971). The best fit gaussian function can account for
47.6 stars amongst a total of 81.3 photometric members. The binary fraction of 
NGC 4609 is therefore 41.5 \% (= $100 \times ( 81.3 - 47.6 ) / 81.3$).
Two factors may affect this fraction. Contamination by field stars
may cause the binary fraction to be overestimated because field stars seem to be
relatively older than NGC 4609 and are therefore slightly brighter. Another is
the effect of binary systems with a large mass ratio. In that case, the
light contribution from the faint companion is insignificant and
we may underestimate the binary fraction. At this point we cannot estimate
which effect affects the results more.

Fig. \ref{cmd} shows the CMDs of the observed region. The blue members 
as well as photometric members of each cluster are marked with different
symbols. There are several photometric members with red $J-H$ colors.
Most of these are due to the photometric errors in 2MASS (upper limit in $H$).
Only two stars (ID 160 = 2MASS J12410653-6304343 and ID 6987 = 2MASS 
J12440630-6303586) seem to have normal optical colors and red near-IR colors.
Their nature should be checked by spectroscopic observations.

\subsubsection{Hogg 15}

The radial variation of reddening in the Hogg 15 region is shown in the right
panel of Fig. \ref{redmap}. The amount of reddening near the cluster center is
small, but increases abruptly outside $r \approx 2'$. A similar variation
pattern can be seen in the starburst-type young open cluster NGC 3603
(see Fig. 5 of Sung \& Bessell 2004). Such a variation may be caused by 
strong winds from young massive stars.

The distance modulus of Hogg 15 is estimated to be 12.50 $\pm$ 0.15 mag
($d =$ 3.2 $\pm$ 0.2 kpc) from the ($M_V$, $B-V$) and ($M_V$, $U-B$) relations
for the ZAMS stars and the reddening-corrected magnitude and colors of young blue 
stars in Hogg 15. This value is somewhat smaller than the distance obtained
by Moffat (1974 - $d =$ 4.2 $\pm$ 0.4 kpc) and van der Hucht (2001 - $d =$ 
3.80 kpc). Sagar et al. (2001) estimated the distance to Hogg 15 as 3.0 
$\pm$ 0.3 kpc, while Piatti \& Clari\'a (2001) obtained a much smaller
value ($d =$ 2.6 $\pm$ 0.08 kpc).
We have selected photometric members of the Hogg 15 group in the restricted area
($\Delta \alpha =$ -5$'$ -- 10$'$ and $\Delta \delta=$ -5$'$ -- -1$'$) where
we could find highly reddened early-type stars. Only 13 and 3 photometric 
members, respectively, have been selected within and outside the cluster radius.

\section{AGE AND MASS FUNCTION}

\subsection{NGC 4609}

\subsubsection{Age of NGC 4609}

\begin{figure}[!h]
\begin{center}
\epsfxsize=5.6cm \epsfbox{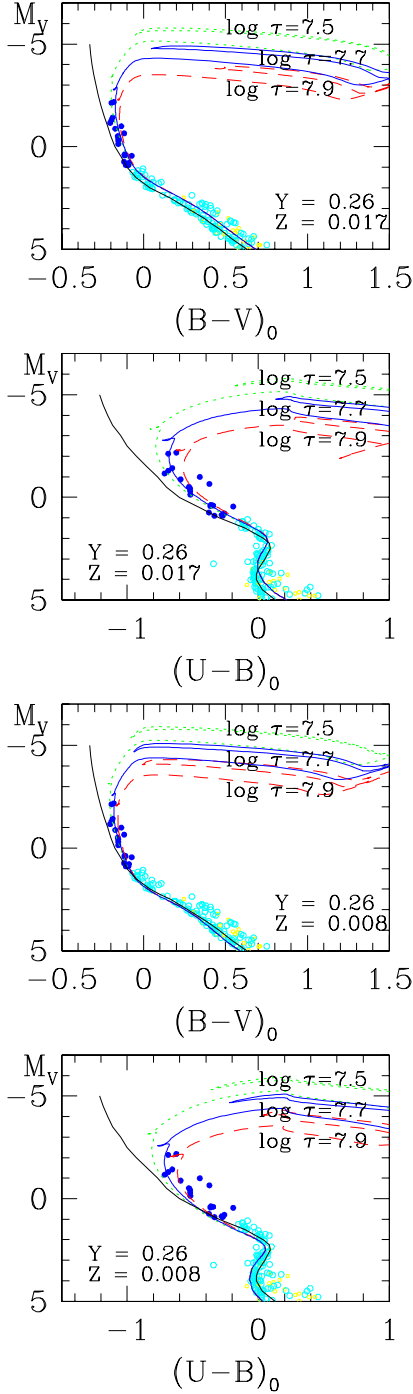}
\end{center}
\caption{The observational H-R diagram of NGC 4609 ($r \leq 6.'5$). 
The superimposed isochrones of $\log age =$ 7.5, 7.7, and 7.9
from Bertelli et al. (2009) are for solar abundances (upper two panels)
and for Z = 0.008 (lower two panels). The thick solid line represents the 
empirical
ZAMS relation. All other symbols are the same as in Fig. \ref{ccd}.
\label{n4609hrd} }
\end{figure}

Feinstein \& Marraco (1971) determined the age of NGC 4609 ($age = 6 \times
10^7$ yrs). In order to determine the age of NGC 4609 using the most recent
isochrones we have drawn the observational Hertzsprung-Russell (H-R) diagram of
NGC 4609 in Fig. \ref{n4609hrd}. Superimposed are the most recent isochrones 
from the Padua group (Bertelli et al. 2009) for solar abundances (upper two panels)
and for Z = 0.008 (lower two panels). Clari\'a et al. (1989) observed a red 
giant candidate in NGC 4609 - No 43 of Feinstein \& Marraco (1971) - in $UBV$, 
DDO, and Washington systems, but they did not arrive at any conclusion
regarding the metallicity of NGC 4609.
Although the star is outside the cluster boundary, they assumed it to be
a member of NGC 4609 from its $E(B-V)$ and spectral type estimated from 
DDO photometry. As mentioned in Section 3.3.1, the photometric characteristics
of members and field stars around NGC 4609 are very similar, and so
the reddening $E(B-V)$ by itself cannot be a decisive membership criterion.

As mentioned in Cheon et al. (2010), 
the ZAMS of the solar abundance models of Padua group is brighter than 
the empirical one by about 0.3 mag. In spite of the difference in abundances
and/or ZAMS relations, the age of NGC 4609 is well fitted to the isochrone
of age $5 \times 10^7$ yrs. As there is only a small difference in the age from
both diagrams we therefore
adopt the age of NGC 4609 as $\log age = 7.7 \pm 0.1$.

\subsubsection{Luminosity and Mass Function of NGC 4609}

\begin{figure}[h]
\begin{center}
\epsfxsize=7.0cm \epsfbox{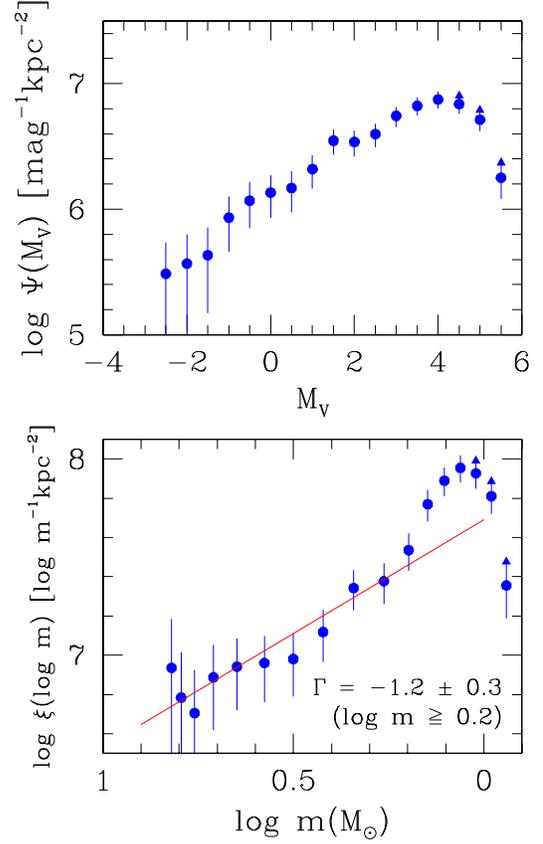}
\end{center}
\caption{The luminosity and mass function of NGC 4609.
\label{n4609lf} }
\end{figure}

The slope between the absolute magnitude and mass, i.e. ${d M_V} \over 
{d \log m}$, is relatively large for intermediate- 
or low-mass stars. It is easy to estimate the mass of a star in intermediate-age
open clusters reliably from its absolute magnitude.
We first calculate the luminosity function (LF) of the photometric members
of NGC 4609 after subtracting the contribution of field stars.
The LF of NGC 4609 is calculated with the bin size ($\Delta M_V$) of 1 mag.
And we also calculate the LF with the same bin size but shifted by 0.5 mag
to minimize the effect of binning. The LF of NGC 4609 is presented in
the upper panel of Fig. \ref{n4609lf}.

The mass function (MF) of NGC 4609 can be derived from the relation 
$ \xi (\log m) d \log m \equiv \Psi (M_V) d M_V$ and the adopted
mass-luminosity relation. In calculating the MF of NGC 4609, we use
the mass-luminosity relation of the isochrone of $\log age = 7.7$
and Z = 0.008 (Bertelli et al. 2009). The MF obtained here is shown 
in the lower panel of Fig. \ref{n4609lf}. The slope of the MF for
$\log m \geq 0.2$ is -1.2 $\pm$ 0.3. For low-mass stars ($\log m < 0.2$)
some enhancement is readily seen which may be attributable to 
contamination from field stars. As already mentioned in Section 3.3.1,
photometric membership criteria do not work well if the photometric
characteristics of cluster stars and field stars are similar.
For brighter stars photometry is relatively accurate, and so the photometric
membership criteria work well. But for fainter stars the error in
photometry increases rapidly, and therefore we can expect the inevitable 
contamination by field stars.

\subsection{Hogg 15}

\subsubsection{The Age of Hogg 15}

\begin{figure}[!h]
\begin{center}
\epsfxsize=7.0cm \epsfbox{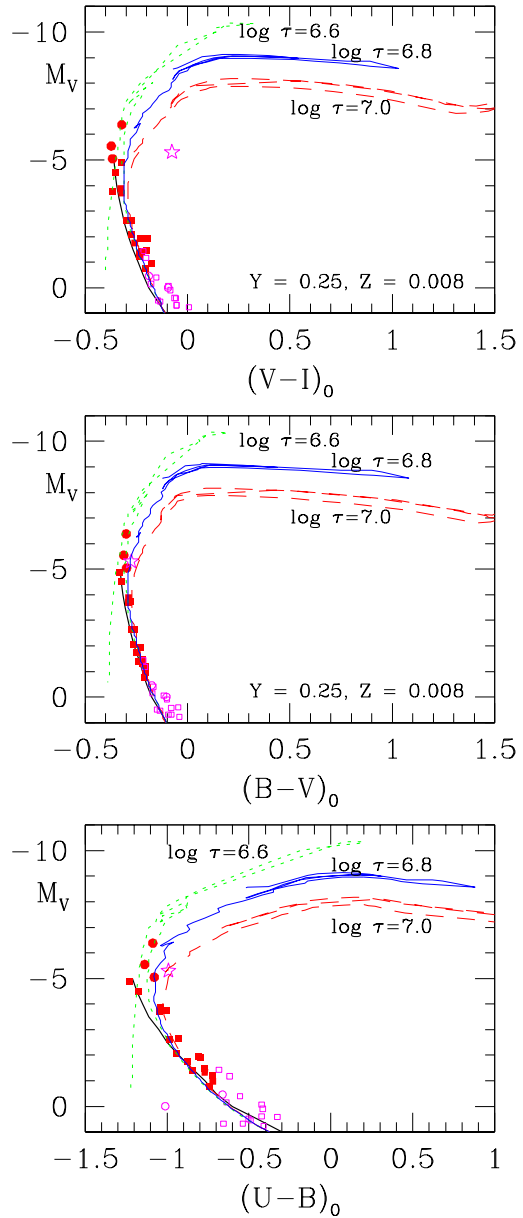}
\end{center}
\caption{The observational H-R diagrams of Hogg 15. The thick solid line 
represents
the empirical ZAMS relation, while the thin dotted, solid, and dashed lines are
the isochrones of $\log age =$ 6.6, 6.8, and 7.0, respectively for Z = 0.008
from Bertelli et al. (1994). Squares and circles denote the member stars
within and outside of Hogg 15, respectively. Filled and open symbols represent
the young blue stars and less bright photometric members, respectively.
The star symbol represents the position of WR 47.  \label{ho15hrd} }
\end{figure}

Moffat (1974) estimated the age of Hogg 15 as $\lesssim 8 \times 10^6$ yrs
probably from the earliest photometric spectral type ($b0$) of stars in Hogg 15.
Later Sagar et al. (2001) determined the age as  6 $\pm$ 2 Myr, but
Piatti \& Clari\'a (2001) derived a much older age ($\sim$ 300 Myr).
We constructed the observational H-R diagram of Hogg 15 in Fig. \ref{ho15hrd}.
As Bertelli et al. (2009) do not provide isochrones for $\log age < 7.0$,
we superimpose the isochrones of a younger age from Bertelli et al. (1994).
For OB stars, metallicity hardly affects the optical colors; we use
the isochrones for Z = 0.008.

Filled and open symbols in Fig. \ref{ho15hrd} represent stars within and 
outside the radius of Hogg 15. The stars in Hogg 15 are very close to the 
empirical ZAMS, and younger than the youngest isochrone ($\log
age$ = 6.6). The young stars outside the radius of Hogg 15 are on the
isochrone of $\log age =$ 6.6. Although WR 47 (star symbol in Fig. 
\ref{ho15hrd}) seems to be older than the others, its position in Fig.
\ref{ho15hrd} cannot indicate the age of Hogg 15 because its colors may 
be affected by the emission lines of WN6 stars. The age of Hogg 15 is 
younger than or similar to 4 Myr ($\log age \lesssim 6.6$).

\subsubsection{The Initial Mass Function of Hogg 15}

\begin{figure}[h]
\begin{center}
\epsfxsize=7.0cm \epsfbox{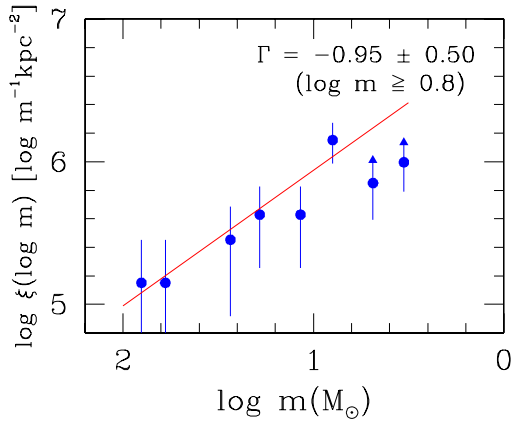}
\end{center}
\caption{The initial mass function of Hogg 15.
\label{ho15imf} }
\end{figure}

The mass of individual members in Hogg 15 is estimated using the Padua
isochrone of $\log age =$ 6.6. The mass of WR 47 (WN6 + O5:V) is
from Lamontagne et al. (1996). Although the observed mass of WN6 stars
is 51 M$_\odot$, its initial mass is much larger than its companion (O5:V
- 60 M$_\odot$). We tentatively assume, therefore, that the mass of the WN6 
star is 80 M$_\odot$ (see Section 5.1 for more discussion). 

The initial mass function 
(IMF) of Hogg 15 is derived and presented in Fig. \ref{ho15imf}. For Hogg 15
we do not attempt to subtract the contribution of field stars because
the photometric members are blue stars and such blue stars are not
found in the field.\footnote{The young blue stars outside the radius of Hogg 15
may be members of an unclassified association around Hogg 15.} The slope of
the IMF is -0.95 $\pm$ 0.50 for $\log m \geq 0.8$, which is somewhat 
shallower than normal. The large error is caused by the small number of 
member stars.

\section{DISCUSSION}

\subsection{Age and Initial Mass of the WN6 Stars in Hogg 15}

\begin{figure}[h]
\begin{center}
\epsfxsize=7.0cm \epsfbox{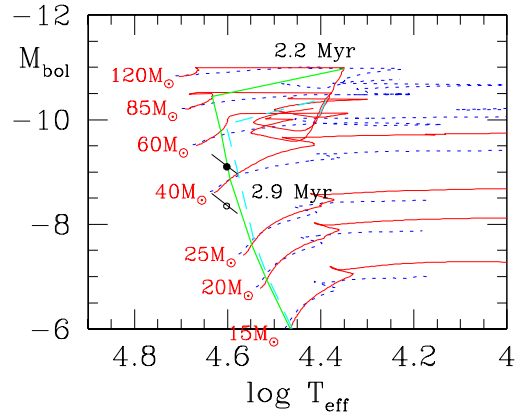}
\end{center}
\caption{The H-R diagram of the O5:V star in WR 47. Thin solid and dotted 
lines with mass to the left are stellar evolution models from Meynet \& Maeder
(2003) with and without stellar rotation, respectively. Thick solid and dashed
lines represent the isochrones from evolution models with stellar rotation
($v_{\rm rot} = 300$ km s$^{-1}$) of age 2.2 and 2.9 Myr, respectively.
Open and filled circles are the position of the O5:V star for two cases. 
See text for details.  \label{wr47} }
\end{figure}

To estimate the initial mass of the WN6 star in WR 47, we have drawn 
the upper part of the H-R diagram in Fig. \ref{wr47}. 
The theoretical stellar evolution models with rotation constructed by
the Geneva group (Meynet \& Maeder 2003) are superimposed.
The absolute visual magnitude of WR 47 is -5.30 mag if we apply the $E(B-V)$ 
derived from Fig. \ref{ccd} by assuming the broad-band color of 
the system is dominated by light from the normal 
companion O5:V star. We estimate the absolute bolometric magnitude of O5:V in
two ways - (1) both O5:V and WN6 equally contribute to the total light, and
(2) the light from O5:V star dominates in the WR 47 system. The open circle
in Fig. \ref{wr47} is for case (1), and the filled circle is for
case (2). The error bar is for a 1 subclass error in the spectral type of the 
O5:V star, i.e. O4V -- O6V. The mass and age of the O5:V star are, \\ 
case (1): ($\sim$32 M$_\odot$, $\sim$1.7 Myr) -- ($\sim$40 M$_\odot$, 
$\sim$ ZAMS) \\
case (2): ($\sim$40 M$_\odot$, $\sim$2.9 Myr) -- ($\sim$52 M$_\odot$,  
$\sim$1.3 Myr)

The most massive star in the Geneva models (120 M$_\odot$, 
total lifetime : 3.55 Myr) becomes 
an early phase WNL at about 2.15 Myr ($\sim$ 86 M$_\odot$), 
a WNE at about 3.19 Myr ($\sim$ 23.5 M$_\odot$), and a WC at 
about 3.26 Myr ($\sim$ 20.5 M$_\odot$). And a 85 M$_\odot$ star (total 
lifetime : 4.09 Myr) becomes WNL, WNE, WN/WC, and WC phase at 2.68 Myr
($\sim$ 64.6 M$_\odot$), 3.75 Myr ($\sim$ 20.8 M$_\odot$), 3.88 Myr 
($\sim$ 17.3 M$_\odot$), and 3.885 Myr ($\sim$ 17.1 M$_\odot$), respectively.
The observational mass ratio ($q_{WR}$) is 0.85. 

The maximum age of case (1) is 1.7 Myr which is untenable because even a 120 
M$_\odot$ star cannot evolve into the WNL phase in this time.
There are three possibilities for case (2). 
The most massive case ($\sim$ 52 M$_\odot$ for the companion) is also  
unacceptable because of the age. At the age of 2.2 Myr for O5 star, a 120 
M$_\odot$ star would have just evolved into the WNL phase, and have a mass of
about 83.7 M$_\odot$, which is much larger than the mass of the WN6 star 
($q_{WR} \cdot$ M$_{O5:V} \approx$ 38 M$_\odot$).

The only remaining case is the case of (M$_{O5:V} \sim$40 M$_\odot$, $age \sim$2.9 
Myr).  The mass of the WN6 is $q_{WR} \cdot$ M$_{O5:V} \approx$ 34 M$_\odot$, and
the evolutionary mass of m$_{init}$ = 120 M$_\odot$ at 2.9 Myr is 36.2 
M$_\odot$. These two values are very similar. In addition, the maximum mass loss
from the current mass loss rate (Moffat et al. 1990) is about 87 M$_\odot$ 
(= $\dot{M} \cdot \tau_{age} = 3 \times 10^{-5} $M$_\odot yr^{-1} \cdot 
2.9 \times 10^6 yr$). This value is consistent with the mass obtained above
(M$_{WN6} >$ 33 M$_\odot$ = 120 M$_\odot$ - 87 M$_\odot$). We obtain a much
smaller mass for the WR 47 system (M$_{WR ~47} = 74$ M$_\odot$ from this)
compared to the 111 M$_\odot$
in Lamontagne et al. (1996). Niemela et al. (1980) derived mass ratios and
masses for several cases. But for all cases the total mass of the system
was larger than 82 M$_\odot$.

In this calculation we have implicitly assumed that the mass ratio obtained 
from observations is correct, that the mass of individual components may be 
incorrect, that the WR system is coeval, and that the stellar evolution models 
are correct. In addition, we assumed no mass transfer between components. 
The above result may be meaningless if even one of the above assumptions 
are invalid. Another problem that needs to
be solved is the bolometric correction (BC) scale of WR stars. 
The M$_{bol}$ of m$_{init}$ = 120 M$_\odot$ at 2.9 Myr is about -10.1 mag. 
The absolute visual magnitude of the WN6 star will be -4.7 mag if we apply BC = 
-5.4 mag (Nugis \& Lamers 2000). In this case the light from the WN6 star 
dominates that from the O5:V star. 

The predictions from stellar evolution models and observations are 
inconsistent for the WR 47 system. As the absolute parameters for
each star in the system are very important in the observational tests
of the formation and evolution of massive stars, more observational studies
should be made to clarify the evolutionary status of WR 47.

\section{SUMMARY}

$UBVI$ CCD photometry data has been obtained for the intermediate-age open cluster
NGC 4609 and the young open cluster Hogg 15. The results obtained may be
summarized as follows.

1. A nearly normal reddening law ($R_V = 3.2$) is determined for the observed 
region from the color excess ratios of highly reddened early type stars in the
observed region.

2. The distance and median value of interstellar reddening of NGC 4609 are 
10.60 $\pm$ 0.08 mag ($d = 1.32 \pm 0.05$ kpc) and 0.37 $\pm$ 0.03 mag, 
respectively. The corresponding parameters of Hogg 15 are 12.5 $\pm$ 0.15 
mag ($d = 3.2 \pm 0.2$ kpc) and 1.13 $\pm$ 0.11 mag.

3. The radius of NGC 4609 is estimated to be $r = 6.'5$ from the surface density
variation of bright B type stars in the observed region. That of Hogg 15 is
also determined to be $r = 3.'5$.
Hogg 15 is found to be a cluster with a core and a halo surrounding the core.

4. The WR star HDE 311884 is in the halo of Hogg 15. Its reddening,
brightness and age of Hogg 15 strongly imply that the star is 
a member of Hogg 15.

5. Using the photometric membership selection criteria, we select cluster
members down to $V = 17$ mag. 

6. The age and mass function of NGC 4609 and Hogg 15 were derived. The age of
NGC 4609 is $\log age = 7.7 \pm 0.1$, while that of Hogg 15 is $\log
age \lesssim 6.6$.  The slope of the mass function of NGC 4609 is nearly normal,
but in the low-mass regime the mass function seems to be affected by field
interlopers. On the other hand, the IMF slope of Hogg 15 is slightly shallow
($\Gamma = -0.95 \pm 0.5$).

7. We have also discussed the mass and evolutionary status of
the double-line binary system WR 47 (WN6 + O5:V).

\acknowledgments

H. S. acknowledges the support of the National Research Foundation of Korea
(NRF) to the Astrophysical Research Center for the Structure and Evolution 
of the Cosmos (ARCSEC$''$) at Sejong University (NRF No. 2009-0062865).

\end{document}